\documentclass[12pt]{article}
\usepackage{subfigure}
\usepackage{graphicx}
\usepackage{cite}
\usepackage{amsmath}

\begin{document}
\title{Fish schooling as a basis for vertical axis wind turbine farm design\footnote{The technology described in this paper is protected under both US and international pending patents filed by the California Institute of Technology.}}
\author{Robert W Whittlesey$^1$, Sebastian Liska$^1$ and John O Dabiri$^{1,2}$\\
\\
$^1$ Graduate Aeronautical Laboratories,\\ California Institute of Technology, Pasadena CA 91125, USA\\
$^2$ Option in Bioengineering, \\California Institute of Technology, Pasadena CA 91125, USA}
\maketitle

\begin{abstract}

Most wind farms consist of horizontal axis wind turbines (HAWTs) due to the high power coefficient (mechanical power output divided by the power of the free-stream air through the turbine cross-sectional area) of an isolated turbine. However when in close proximity to neighbouring turbines, HAWTs suffer from a reduced power coefficient. In contrast, previous research on vertical axis wind turbines (VAWTs) suggests that closely-spaced VAWTs may experience only small decreases (or even increases) in an individual turbine's power coefficient when placed in close proximity to neighbours, thus yielding much higher power outputs for a given area of land. A potential flow model of inter-VAWT interactions is developed to investigate the effect of changes in VAWT spatial arrangement on the array performance coefficient, which compares the expected average power coefficient of turbines in an array to a spatially-isolated turbine. A geometric arrangement based on the configuration of shed vortices in the wake of schooling fish is shown to significantly increase the array performance coefficient based upon an array of $16\times16$ wind turbines. Results suggest increases in power output of over one order of magnitude for a given area of land as compared to HAWTs.

\end{abstract}

\section{Introduction}
Technology for extracting energy from the wind has become a topic of increasing political, economic and social interest in the last fifty years~\cite{ackermann2005wind,20by2030}. Nearly all current wind farms use horizontal axis wind turbines (HAWTs) due to their high power coefficient, defined as the mechanical power output of the turbine divided by the power of the free-stream air through the cross-sectional area of the rotor normal to the free-stream direction. However, a major constraint on wind farms is the inter-turbine spacing, since the performance of each HAWT decreases with decreased spacing~\cite{lissaman1979energy,hauBook,liu1983mathematical,barthelmie2007modelling}. When a HAWT extracts energy from the free stream, the speed of the flow behind the turbine is reduced. A well-known aerodynamic result is that the maximum amount of power a rotor can extract is proportional to the cube of the free-stream speed \cite{hauBook}. Horizontal axis wind turbines downstream of other HAWTs have a reduced incoming flow speed and thus are capable of extracting significantly less power than that of a spatially-isolated turbine. Theoretical analysis, computational simulations, and experimental results indicate that decreasing the downstream spacing of a row of multiple HAWTs from ten rotor-diameters to five rotor-diameters can decrease the power generated by roughly 40\%~\cite{jensen1984note,liu1983mathematical,ivanell2005numerical}. Consequently, most modern HAWT wind farms space turbines five to ten rotor-diameters downstream and three or more rotor-diameters laterally~\cite{sorensen2004renewable,hauBook}.

An alternative paradigm for wind energy extraction is found in vertical axis wind turbine (VAWT) designs. Studies have shown that a single spatially-isolated VAWT has a significantly lower power coefficient and costs approximately a third more than a HAWT with comparable power output~\cite{hauBook}. Current literature regarding the performance of VAWT wind farms is scarce when compared to HAWTs, yet some sources indicate that the typical spacing constraints on HAWT wind farms do not apply~\cite{schatzle1980aerodynamic,nguyen1978vortex,rajagopalan1990aerodynamic}. Furthermore, a recent U.S. Patent suggests that aerodynamic interactions between a pair of counter-rotating VAWTs contribute to a higher power coefficient per turbine~\cite{patent}. Previous work by Rajagopalan, Klimas, and Rickerl suggests an increase of up to 4\% in the average turbine power coefficient of an array of VAWTs due to steam-tube contraction (induced higher flow speeds around neighbouring VAWTs)~\cite{rajagopalan1990aerodynamic}. In contrast, Schatzle, Klimas, and Spahr show no change or a mild reduction in power coefficient for a pair of closely-spaced VAWTs~\cite{schatzle1980aerodynamic}. However, neither of these studies considered the possibility of alternating the directions of rotation of neighbouring VAWTs.

For this work, we consider a VAWT array configuration following the arrangement of shed vortices in the wake of schooling fish. These shed vortices form a reversed K\'{a}rm\'{a}n vortex street in the wake of the fish~\cite{fishlikeSwimming}. Previous work by Weihs \cite{weihs} uses a potential flow model to analyze the wake effects of the shed vortices on a school of fish. Motivated by the demonstrated benefits of the reversed K\'{a}rm\'{a}n vortex street on the propulsion of the schooling fish, we apply the same configuration and similar modelling tools to analyze VAWT arrays.

We investigate the inter-turbine spacing effects of VAWT arrays through a semi-empirical model. Models are developed, given in section \ref{model}, to describe a single-VAWT and arrays of VAWTs using parameters from the field measurements. The field measurements of an installed VAWT are presented in section \ref{exper}. The results of fitting the model parameters to the field measurements and evaluating effects of array configuration on power output are given in section \ref{results} and discussed in section \ref{disc}.

\section{Modelling}\label{model}
\subsection{Potential Flow Model}\label{basicModelling}

In the interest of examining the effects of spacing and orientation of VAWTs in an array, a potential flow model is first constructed for a single turbine. The model consists of uniform flow with a dipole and a point vortex superimposed to represent one VAWT. The point vortex accounts for the flow rotation caused by the turbine and the dipole represents flow blockage due to the presence of the turbine in the flow. The complex velocity potential, $W$, and the velocity vector field, $\bf u$, are given by:
\begin{equation}
W = U_\infty z - \underbrace{i \frac{\Gamma}{2 \pi} \log(z) + \mu z^{-1}}_{One\ VAWT}
\label{potModel}
\end{equation}
\begin{equation}
{\bf u} = \Re(dW/dz) \hat{\imath} - \Im(dW/dz) \hat{\jmath}
\label{veloEqn}
\end{equation}
where $U_\infty$ refers to the free-stream speed, $\Gamma$ is the strength of the point vortex, $\mu$ is the strength of the dipole, $\hat{\imath}$ and $\hat{\jmath}$ are the unit vectors in the $x$ and $y$ directions, respectively, and $z = x + i y$ ($x$ and $y$ are real numbers). To create the velocity potential corresponding to an array of $K$ VAWTs, the model in equation \ref{potModel} is summed over the VAWT positions in the array,
\begin{equation}
W = U_\infty z + \displaystyle\sum_{k = 0}^K\underbrace{\Biggl [ -i \frac{\Gamma}{2\pi} \log \left(  z - z_k \right) + \mu \left( z - z_k \right )^{-1} \Biggr ]}_{kth\ VAWT}
\label{potArray}
\end{equation}
where $z_k$ corresponds to the location of the $k$th VAWT. The velocity field is obtained using equation \ref{veloEqn}. To account for a velocity-deficit in each VAWT wake, the velocity-deficit curves given by Hau \cite{hauBook} are applied:
\begin{equation}
{\bf u^{*}}(z) =\Bigl(1-\xi_{w}(z)\Bigr)\ {\bf u}(z)
\label{wakeEqn}
\end{equation}
where $\bf u^*$ is the resulting velocity vector after accounting for wake-effects and $\xi_{w}(z)$ is a spatial function representing the velocity deficit. The function $\xi_w(z)$ is a normal probability density function (PDF) in the lateral direction and a beta PDF in the downstream direction. The term $\Bigl(1-\xi_{w}(z)\Bigr)$ adjusts the magnitude of the downstream velocity to reflect the velocity deficit in the wake. The wake extends approximately six rotor-diameters downstream and 20$^\circ$ bilaterally from the free-stream wind direction. This velocity deficit is included for each VAWT in the array. Accounting for the wake in this manner implies that ${\bf u^*}$ is no longer a solution to the potential flow equations.

In order to evaluate the expected power produced by a turbine in an array, we define a parameter representative of the power output. By considering a simple drag-based VAWT, the tangential force on each turbine blade is $\sim u_t^2$, where $u_t$ is the velocity tangent to the blade motion. Furthermore, we assume that angular velocity of the rotor is $\sim u_t$. Putting these together, the model for the power output of a single, spatially-isolated VAWT can be represented by:
\begin{equation}
P_{iso} = C_{iso} \frac{\rho R_C}{2\pi}\oint_{C} \left( \mathbf{u^*} \cdot \hat{\mathbf{s}} \right)^3 ds
\label{powerPara}
\end{equation}
where $P_{iso}$ is the power output of the turbine, $\rho$ is the density of air, $C_{iso}$ is a non-dimensional parameter specific to a VAWT design, $R_C$ is the radius of the wind turbine, $\hat{\mathbf s}$ is the unit vector tangent to the contour of integration, and $C$ is the contour of integration that is a circle with radius $R_C$ centered on the VAWT in the direction of rotation. The non-dimensional parameter $C_{iso}$ is included to allow for the variety of wind turbine designs and accounts for effects such as blade shape, height of turbine, and other factors that affect wind turbine performance. The same expression for $P_{iso}$ (with a different $C_{iso}$) is obtained for a lift-based VAWT model as the lift is also $\sim u_t^2$ and it is still reasonable to assume the angular velocity of the rotor is $\sim u_t$.

Having defined $P_{iso}$ we introduce the array performance coefficient, $C_{AP}$, to evaluate the effects of closely spaced turbines in an array:
\begin{equation}
C_{AP} = \displaystyle \frac{\overline{P_{array}}}{P_{iso}}
\label{CAP}
\end{equation}
where $\overline{P_{array}}$ is the average performance parameter for an array. It is implicitly assumed that all turbines in an array are homogeneous (i.e. same $C_{iso}$). For $C_{AP} < 1$, the average VAWT in the array performs worse than an equivalent, spatially-isolated VAWT. However, if $C_{AP} > 1$, the model suggests that the array configuration provides beneficial, inter-VAWT, aerodynamic interactions and increases the average power output of the array.

Although comparing $C_{AP}$ is a means for understanding the flow physics, as a means to compare the performance of an array configuration in terms of power output per unit land area occupied by the VAWT farm, a power density parameter, $C_{PD}$, is defined:
\begin{equation}
C_{PD} = \left (\frac{P_{array}}{A_{array}}\right)\Bigg/\left(\frac{P_{iso}}{A_{iso}} \right)= K \cdot C_{AP} \frac{A_{iso}}{A_{array}}
\end{equation}
where $A_{array}$ is the land area of the entire wind farm, $A_{iso}$ is the land area of a single turbine, taken to be arbitrarily $D^2$, and $K$ is the number of turbines in the array. This parameter is most useful in the practical design of a wind farm and determining the appropriate spacing for maximizing the power output per a given area.

\subsection{BioInspiration}
\begin{figure}
\centering
\subfigure[Configuration of shed vortices in the wake of schooling fish. Adapted from Weihs \cite{weihs}. Coordinate system moves with the school, the fish are headed in the negative $\hat{\imath}$ direction.]{\label{weihsFish}\includegraphics{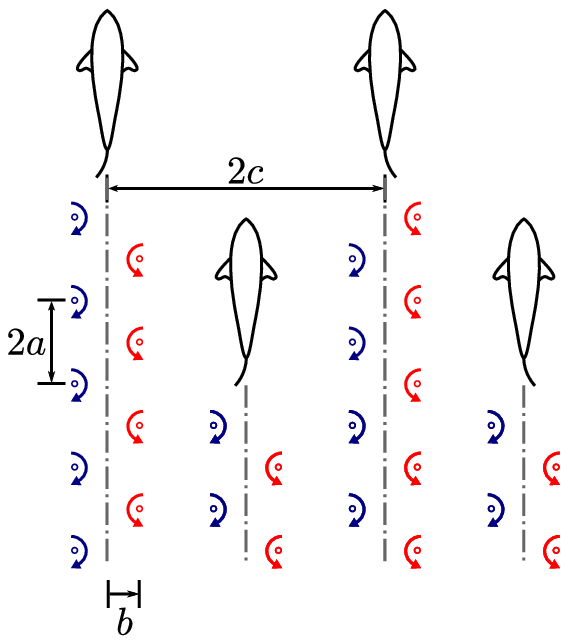}}
\hspace{0.2cm}
\subfigure[Configuration for the VAWT wind farm based upon shed vortices in the wake of schooling fish. $U_\infty$ is in the positive $\hat{\imath}$ direction.]{\label{turbineFish}\includegraphics{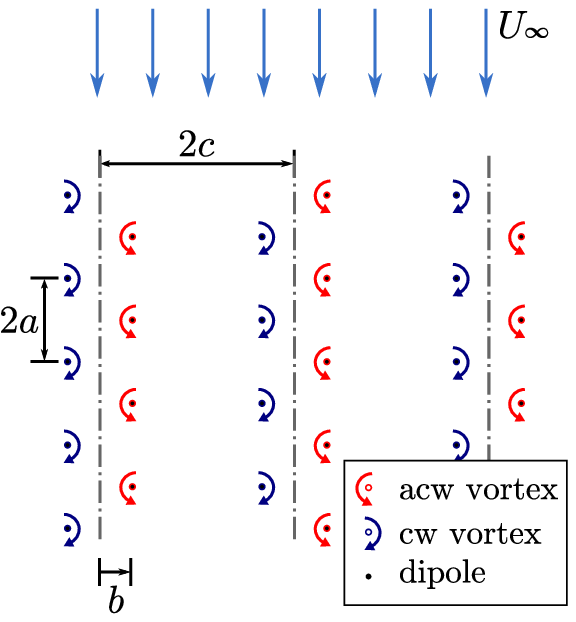}}
\caption{Diagrams comparing the model using shed vortices from schooling fish by Weihs \cite{weihs} with the arrangement of VAWTs used in the current study.}
\label{fishArray}
\end{figure}

A mathematical model similar to that described in the previous section is used by Weihs \cite{weihs} to study the wake effects of fish schooling arrangements on fish propulsion. Weihs' potential flow model uses uniform flow and superimposes point-vortices at the location of shed vortices from fish as they swim, depicted in figure \ref{weihsFish}. The location of these shed vortices can be described in the complex plane by
\begin{align}
A_{m,n}^{1+} &= -a/2 + 2 n  a + i(4 m c + b)\\
A_{m,n}^{2+} &= a/2 +  2na + i((4m + 2) c + b) 
\label{posVort}
\end{align}
\begin{align}
A_{m,n}^{1-} &= a/2  + 2 n  a + i(4 m c -b)\\
A_{m,n}^{2-} &= -a/2  + 2 n  a + i((4m + 2) c -b)
\label{negVort}
\end{align}
where $A_{m,n}^{1+}$ and $A_{m,n}^{2+}$ correspond to positive (anticlockwise rotating) vortices, $A_{m,n}^{1-}$ and $A_{m,n}^{2-}$ correspond to negative (clockwise rotating) vortices, $2a$ is the downstream distance between two vortices of the same line, $2b$ is the lateral distance between the two lines of shed vortices from a given fish, $2c$ is the distance between two adjacent fish, and $m$ and $n$ are integers. Fish in a school typically swim with $b/a \approx 0.3$, depending on species, and $c/b \geq$ 2 \cite{weihs}. Incorporating equation \ref{posVort} and equation \ref{negVort} into equation \ref{potArray}, the resulting velocity potential for an array of VAWTs following this particular arrangement is
\begin{eqnarray}
W = U_{\infty} z & - i \frac{\Gamma}{2 \pi} \sum_{m = 1}^{M} \sum_{n = 1}^{N} \Bigl[&\log(z-A_{m,n}^{1+}) + \log(z-A_{m,n}^{2+})  \nonumber\\
& &- \log(z - A_{m,n}^{1-}) - \log(z-A_{m,n}^{2-}) \Bigr] \nonumber\\
& \quad + \mu \sum_{m = 1}^{M} \sum_{n = 1}^{N} \Bigl[& (z-A_{m,n}^{1+})^{-1} +  (z-A_{m,n}^{2+})^{-1} \nonumber\\
& &  + (z-A_{m,n}^{1-})^{-1} + (z-A_{m,n}^{2-})^{-1}] \label{fishArrayEqn}
\end{eqnarray}

\subsection{Numerical Simulation}\label{numSim}

 equation \ref{fishArrayEqn} is evaluated numerically for a VAWT array configuration where $M~=~N~=~8$, corresponding to an $16\times16$ array of turbines (256 turbines total). The values of $U_\infty$, $\Gamma$, and $\mu$ are obtained from the field measurements, as described in section \ref{dataAssim}. The model is interrogated for the range of parameters: $1D\leq a\leq8D$, -$1.5D \leq b \leq 1.5D$, and $c=2D$, where $D$ corresponds to the turbine diameter. The range of $a$ is chosen to extend from turbines immediately downstream of each other to where the turbines have sufficiently cleared the downstream wake of each other. The range of $b$ is chosen to avoid overlapping turbines.

\section{Field Measurements}\label{exper}

Field measurements of a Savonius-type VAWT located in Glendora, CA were collected in spring 2009 in order to generate input data for the potential flow model. The turbine was a prototype model with rotor radius of $\approx 0.75$ m and rotor height of $\approx 2$ m.
In order to characterize the flow around the VAWT, two cup-type flow anemometers were used. A reference anemometer consisted of an adjustable-length pole with an anemometer (InSpeed Vortex, with a range of 1.4 - 55.6 m s$^{-1}$ and $\pm0.05$ m s$^{-1}$ accuracy) mounted on the end and connected to a dedicated computer that recorded the wind speed every second. The pole was fixed to the ground using guy wires for untended operation. The data from the reference anemometer was averaged over the entire visit to produce the free-stream speed during the field visit.

The second anemometer was mounted to a second pole that was manually positioned at 18 locations around the VAWT, as indicated in figure \ref{vectorsWithModel}. In addition to measuring wind speed, the second anemometer contained a wind vane (InSpeed E-Vane) that measured the direction of the incoming wind ($\pm1^\circ$ accuracy). The wind vane was wired to a computer data acquisition system (Measurement Computing USB-1208FS) which recorded the voltage output of the wind vane at 100 Hz. This was then down-sampled through averaging the 100 measurements in one-second to obtain 1 Hz recordings that synchronized with the wind speed data. The anemometer recorded wind data at each station for approximately three minutes. For our field measurements, both instrument setups were extended to the mid-height of the VAWT and were adjusted according to the underlying terrain ($\approx 6$~m). At the end of the measurements, the mobile instrument setup was moved near the reference setup in order to obtain the free-stream direction, which was assumed constant during the one-hour measuring period. This is a reasonable assumption as local weather stations report only westerly wind during our measurement time period. Data from both anemometers was synchronized for post-processed based on the recorded time-stamps. 

\subsection{Data Assimilation} \label{dataAssim}
The model requires the calculation of three parameters to describe a single, spatially-isolated VAWT: $U_\infty$, $\Gamma$, and $\mu$. To obtain $U_\infty$ from the experimental measurements, the average free-stream speed from the reference anemometer is used. Using this measured free-stream velocity, the values of $\Gamma$ and $\mu$ are determined from a least-squares fit of a single-turbine model (one point vortex and one dipole) to the measured VAWT flow field.

\section{Results}\label{results}

\begin{figure}
\centering
\includegraphics{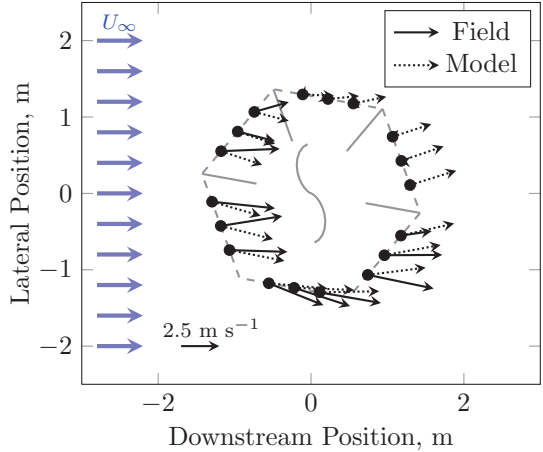}
\caption{The measured flow field from the field visit to the VAWT along with the flow field calculated from equation \ref{potModel}, using $U_\infty~=~3.00$~m~s$^{-1}$, $\Gamma$~=~7.41~m$^2$~s$^{-1}$, and $\mu = 0.00$~m$^3$~s$^{-1}$. The solid vectors indicate the time-averaged experimental velocities. The model's least-squares-fit velocities are superimposed as the dotted vectors. Thin lines correspond to the external frame and rotor of the wind turbine.}
\label{vectorsWithModel}
\end{figure}

The measured flow field from the field visit to the VAWT along with the flow field calculated from equation \ref{potModel}, using the model parameters described in section \ref{dataAssim}, are given in figure \ref{vectorsWithModel}. The measured average free-stream speed, $U_\infty$, is 3.00 m s$^{-1}$ and appears to be well-described by a normal distribution having a standard deviation of 0.86 m s$^{-1}$. The fitting of the measured flow field to the single-VAWT model, as given in equation \ref{potModel}, by least-squared difference yielded a unimodal minimum within the parameter space of $\Gamma$ and $\mu$. The measured results are well-represented by a range of values given by $\Gamma$=7.41$\pm$3.82~m$^2$~s$^{-1}$ and $\mu$ = 0.00$\pm$0.77~m$^3$~s$^{-1}$. Despite the range of $U_\infty$, $\Gamma$ and $\mu$, in this letter we only present results using $U_\infty~=~3.00$~m~s$^{-1}$, $\Gamma$~=~7.41~m$^2$~s$^{-1}$, and $\mu = 0.00$~m$^3$~s$^{-1}$.

\begin{figure}
\centering
\subfigure[$C_{AP}$ for the fish-schooling VAWT-array model.]{\includegraphics{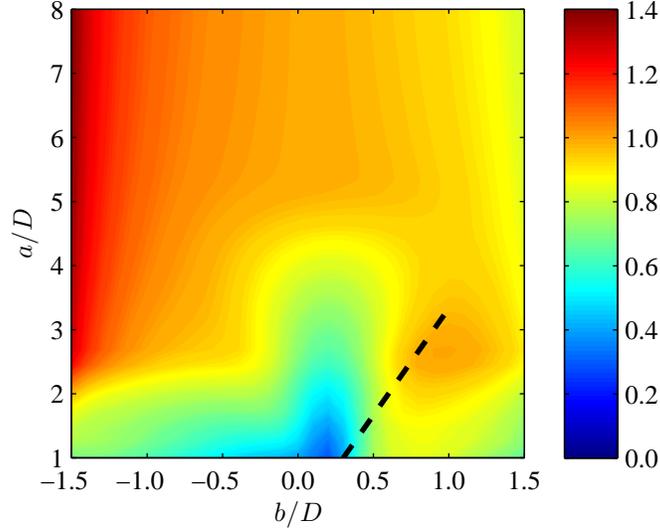}\label{fishyCAP}}
\subfigure[$C_{PD}$ for the fish-schooling VAWT-array model, normalized by the maximum $C_{PD}$.]{\includegraphics{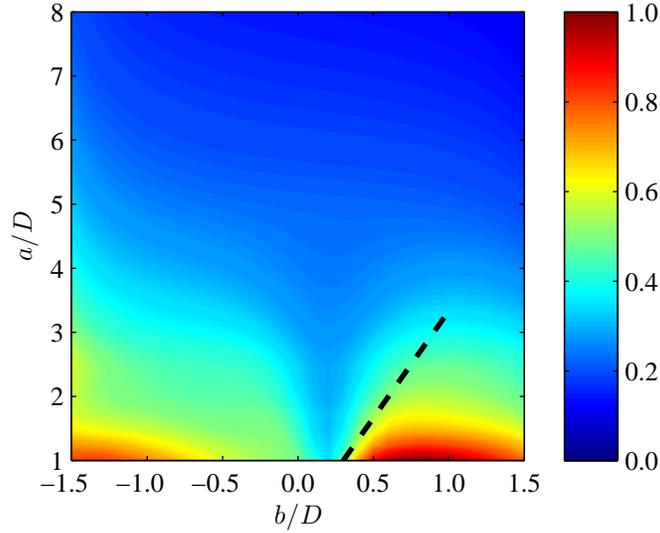}\label{fishyCAPDensity}}
\caption{Calculated $C_{AP}$ and $C_{PD}$ for the fish-schooling VAWT-array model, given by equation \ref{fishArrayEqn}, are shown. Array-spacing parameters, $a$, $b$, and $c$ as shown in figure \ref{turbineFish}, are normalized by the turbine diameter, $D$. For the above figures, $c/D = 2$. The black dashed line corresponds to typical parameters of schooling fish, $b/a = 0.3$ and $c/b \geq 2$.}
\label{fishyCAPandDensity}
\end{figure}

The calculated $C_{AP}$ for the fish-schooling VAWT-array models, given by equation \ref{fishArrayEqn} and using the range of values of $a$ and $b$ described in section \ref{numSim}, is shown in figure \ref{fishyCAP}. There is a clear region for which $C_{AP}$ is greater than unity, corresponding to close-spaced pairs of counter-rotating VAWTs. The maximum $C_{AP}$ in this parameter space suggests the potential for inter-turbine interactions that yield more power output than spatially-isolated turbines.
figure \ref{fishyCAPDensity} shows the normalized power density for the fish-schooling VAWT-array models. Comparing figures \ref{fishyCAP} and \ref{fishyCAPDensity} shows that the optimal array configuration in terms of power extracted per unit land area does not coincide with the largest $C_{AP}$ values, which implies improving the average performance of individual turbines might not be the desired goal of wind farm design.
In figure \ref{fishyCAPandDensity}, the dashed lines show the permissible parameter values for schooling fish. As seen in figure \ref{fishyCAP}, this region is below unity in the $C_{AP}$ results. Array configurations of turbines with $4\times4$ and $32\times32$ turbines were also explored and show results similar to the presented result.

\section{Discussion}\label{disc}

A qualitative comparison between the flow fields from the VAWT model and the collected field data, as shown in figure \ref{vectorsWithModel}, demonstrates that the current single-VAWT model captures the dominant flow characteristics for a single turbine. This verifies the model's ability to reduce a complicated flow field to a three-parameter model. However from the discrepancy between velocity vectors of the single-VAWT model and collected field data in figure \ref{vectorsWithModel}, it is evident that the wake of a spatially-isolated turbine is not accurately captured. These discrepancies motivated the introduction of an empirically-based wake from HAWTs equation \ref{wakeEqn}, which is considered to be an appropriate means of accounting for an expected velocity deficit in the wake when calculating $C_{AP}$.

Measurements were limited to the area immediately surrounding the VAWT. Since the dominant flow effects of the VAWT are most pronounced in the closer locations, the vortex and dipole contributions to the velocity field drop off as $r^{-1}$ and $r^{-2}$, respectively, where $r$ is the distance from the vortex or dipole center. This explains our model's ability to capture the measured flow field despite the aforementioned limitation. Furthermore, measurements farther from the VAWT would likely reflect ambient fluctuations to a greater degree more than VAWT aerodynamics.

\begin{figure}
\centering
\includegraphics[width=5.0in]{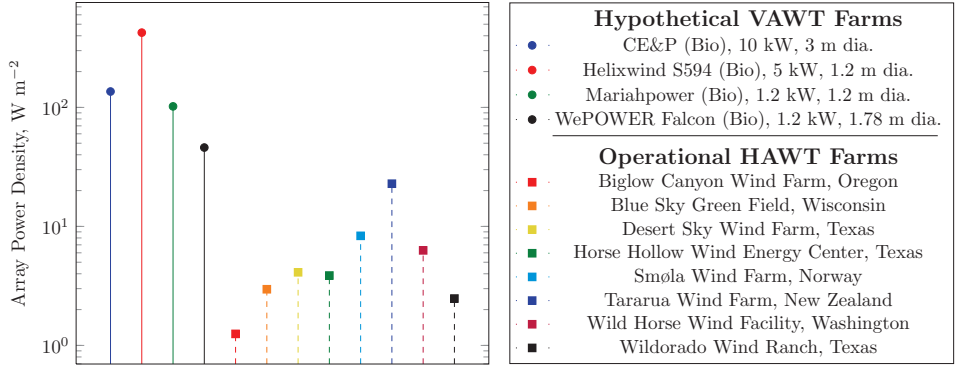}
\caption{Log plot showing the expected array power density, total array power output over array area, of VAWT and HAWT wind farms. Hypothetical VAWT wind farms using geometry based on fish schooling ($a$=$1.2D$, $b$=$0.4D$, $c$=$2D$, $C_{AP}$=0.61) are shown as the circles with solid lines. Geometry chosen based on high $C_{PD}$. A sampling of operational HAWT wind farms are shown as squares with dashed lines. There is a significant increase of over one order of magnitude in the array power density for VAWTs compared to HAWTs.}
\label{landUse}
\end{figure}

In order to understand the practical implications of this work for wind farm design, a comparison of the array power density (total array power output over array area) between VAWTs and HAWTs is made, as given by $C_{PD}$. For this analysis, we assume that the parameters, $\Gamma$ and $\mu$, are the same for all VAWT designs. Additionally, we restrict ourselves to the parameter space representative of schooling fish, and chose the array configuration that maximizes $C_{PD}$ following this constraint based on figure \ref{fishyCAPDensity}. figure \ref{landUse} demonstrates that significantly higher power densities are obtained with VAWT farms compared to operational HAWT farms, with increases of over one order of magnitude. Although close spacing of VAWTs potentially leads to higher power outputs of the individual turbines (regions of $C_{AP}>1$ in figure \ref{fishyCAP}), the major advantage of VAWT wind farms is their significantly smaller inter-turbine spacing, compared to HAWT wind farms, without a substantial decrease in performance as demonstrated by figures \ref{fishyCAP} and \ref{fishyCAPDensity}.

The proposed array configuration adapted from the geometry of fish schooling provides an interesting arrangement for study. figure \ref{fishyCAP} indicates decreases in the average turbine power output if turbines are geometrically arranged following the same configuration of shed vortices in the wake of schooling fish. However the increase in array power density of close-spaced arrays overcomes these mild reductions in expected power output, as seen in figure \ref{landUse}. If one allows for a broader range of parameter space than seen in schooling fish, there is the opportunity for significant improvements ($C_{AP}>1$) in the power output of these turbines. This is not entirely surprising due to differences in the criteria of beneficial arrangements. The fish aim to align themselves to optimize their forward propulsion whereas spatial configurations for turbine arrays aim to maximize energy extraction.

While only one case is presented here, that of a $16\times16$ VAWT array, two other finite arrays were explored with dimensions of $4\times4$ and $32\times32$ turbines. These did not show any significant deviation from the case presented, and thus explains their omission. For all turbine array sizes explored, the wind farm configurations corresponding to the highest values of $C_{AP}$, shown in figure \ref{fishyCAP}, consist of pairs of closely-spaced, counter-rotating VAWTs with their adjoining axis perpendicular to the free stream. A streamline analysis shows that stream-tube contraction between pairs of counter-rotating VAWTs, and the concomitant flow acceleration, is primarily responsible for improvements in expected power output. This is consistent with Rajagopalan et al. \cite{rajagopalan1990aerodynamic}, who studied Darrieus-type VAWT arrays and showed a 4\% increase in power coefficient due to the same mechanism. From our results it appears as though vortex interaction is the dominating mechanism for the stream-tube contraction, rather than blockage effects from neighbouring wind turbines. However, with further study, it is possible that this conclusion is dependent on VAWT design or array configuration, thus allowing VAWT designers to consider blockage effects in future design iterations.

\section{Conclusion}

In this work we present the experimental data used to create a potential flow model to describe the flow field around a spatially-isolated VAWT. This single-VAWT model consisted of three parameters: free-stream speed, vortex strength, and dipole strength. A model for an array of VAWTs, based on the single-VAWT model, is developed and an empirical wake is introduced to calculate the expected power output of the array. To evaluate the potential benefits of a particular array configuration, an array performance coefficient, $C_{AP}$, is defined to compare the average power output of an array configuration to that of a spatially-isolated VAWT. Configurations inspired by the arrangement of shed vortices in the wake of schooling fish are used for interrogating $C_{AP}$ values with varied geometric parameters. Results show the possibility of $C_{AP}$ values up to 1.4, for the $16\times16$ VAWT array, when not restricted to parameters corresponding to actual schools of fish. The leading mechanism for this increase is stream-tube contraction and concomitant flow acceleration between counter-rotating turbines in close proximity. However, values of $C_{AP}$ are approximately or less than unity for configurations following the schooling of fish regardless of array dimensions considered. These configurations significantly reduced the land use for VAWT wind farms, resulting in array power density increases of over one order of magnitude compared to operational HAWT wind farms.

Although the current model does not explicitly allow for complicated flow behavior, such as vortex shedding, turbulence, and three-dimensional effects, the current vortex and dipole model yields a flow field sufficient for assessing the performance of VAWTs. Furthermore, the field data is only from a small segment of time (one hour for the particular visit reported) and does not fully describe the variety of conditions in which VAWTs can produce power. It is of interest to conduct further measurements on a broad range of full-scale VAWT designs and over different operating conditions in order to refine the single-VAWT model. Similar measurements and considerations are required to refine the model for arrays of full-scale VAWTs. Such refinements include investigation of the dependence of $\Gamma$ and $\mu$ on VAWT design, wind speed, and surrounding environment (e.g. turbulence and the atmospheric boundary layer). Including these effects may potentially impact the results in terms of expected power output and array power density.

\section*{Acknowledgments}
We thank J Meier and P Abad-Manterola for assisting with the field work. We also thank California Energy \& Power for allowing us the use of their wind turbine. This work has been funded by the National Science Foundation: Energy for Sustainability Program. The technology described in this paper is protected under both US and international pending patents filed by the California Institute of Technology.

\bibliography{WindFarmReferences}
\bibliographystyle{unsrt}

\end{document}